  \input psfig
  \documentstyle[mprocl,epsf]{article}
  
  \def\Journal#1#2#3#4{{#1} {\bf #2}, #3 (#4)}
  
  \def\NCA{{\em Nuovo Cimento} A}

  \def\NPB{{\em Nucl. Phys.} B}
  \def\PLB{{\em Phys. Lett.}  B}
  \def\PRL{\em Phys. Rev. Lett.}
  \def\PRD{{\em Phys. Rev.} D}
  \def\ZPC{{\em Z. Phys.} C}

  \def\be{\begin{equation}}
  \def\ee{\end{equation}}
  \def\bea{\begin{eqnarray}}
  \def\eea{\end{eqnarray}}
  
  \newcommand{\One}{1\kern-4.5pt1}
  \newcommand{\epsfaxhax}[2]{
          \centerline{
            \hspace{-20pt}
            \epsfxsize=160pt
            {\epsfbox{#1}}
            \hspace{-15pt}
            \epsfxsize=160pt
            {\epsfbox{#2}}}
  }
  
\begin{document}
  
  \title{FIXED POINT FOUR-FERMI THEORIES}
  
  \author{ Simon HANDS}
  
  \address{Department of Physics, University of Wales 
  Swansea,\\
  Singleton Park, Swansea SA2 8PP, UK}
  
  %
  
  \maketitle\abstracts{
I review dynamical chiral symmetry breaking in four-fermi models,
including results of Monte Carlo simulations with dynamical fermions.
For $2<d<4$, where the phase transition defines an ultraviolet fixed
point of the renormalisation group, the continuum theory may either
be describable using the large-$N_f$ expansion, as in the case of the
Gross-Neveu model, or be intrinsically non-perturbative, as in the case
of the Thirring model. For $d=4$, the models are trivial and are 
described by a mean field equation of state with logarithmic corrections
to scaling, which may nonetheless define new universality classes
distinct from those of ferromagnetism.}
  
  \section{Introduction}
  
  In this talk I will review recent progress in our 
  understanding of
  four-fermi theories, by which I mean quantum field 
  theories described
  by the Lagrangian density
  \begin{equation}
  {\cal L}=\bar\psi_i(\partial{\!\!\! /}\,+m)\psi_i
  \pm{{g^2}\over{2N_f}}(\bar\psi_i\Gamma\psi_i)^2.
  \end{equation}
  $\Gamma$ denotes a $4\times4$ matrix constructed from 
  Dirac $\gamma$-matrices, and 
  the sign of the interaction is chosen so as to be 
  attractive
  between fermion and anti-fermion.
The index $i$ runs over $N_f$ 
  distinct fermion 
  flavors. I will discuss these models in spacetime 
  dimension 
  $2<d\leq4$, paying most attention to the cases $d=3$ and 
  $d=4$.
  
  What are the motivations for studying such models? Their 
  generic
  behaviour is that dynamical breaking of a chiral symmetry 
  occurs at some
  strong value of the coupling $g^2$, where 
  non-perturbative methods must
  be applied. In this talk I shall discuss and compare 
  three such methods:
  the large-$N_f$ expansion, use of the Schwinger-Dyson 
  equation, and
  lattice Monte Carlo simulation. Other methods, such as 
  variational~\cite{Yee}
  and derivative expansion~\cite{Aoki} approaches, have 
  also been applied.
  The picture which is emerging is that the symmetry 
  breaking transitions
  in these fermionic models define new universality 
  classes, resembling
  qualitatively, but {\sl not\/} quantitatively, those 
  which apply to the
  study of ferromagnetic phase transitions.
  
  Phenomenologically, four-fermi theories were of course 
  originally
  introduced by Fermi to describe the weak 
  interaction~\cite{Fermi},
  and were next applied to dynamical chiral symmetry 
  breaking in the
  strong interaction by Nambu and 
  Jona-Lasinio~\cite{Nambu}. More recently
  they have appeared in discussions of dynamical mass 
  generation in the
  Standard Model, in such scenarios as walking 
  technicolor~\cite{WTC}, 
  and the top-mode standard model in which the Higgs scalar 
  is a 
  $t\bar t$ bound state~\cite{topmode}. In three spacetime 
  dimensions,
  there may be applications to high-$T_c$ 
  superconductivity~\cite{Shank}, for instance
  in describing non-Fermi liquid behaviour in the normal 
  phase~\cite{Mavro}.
  
  \section{The Gross-Neveu Model for $d=3$}
  
  I will begin by discussing the simplest model, the 
  Gross-Neveu (GN)
  model~\cite{GN}, in which most of the important 
  theoretical issues are
  already present. The Lagrangian is
  \begin{equation}
  {\cal L}_{GN}=\bar\psi_i(\partial{\!\!\! /}\,+m)\psi_i
  -{{g^2}\over{2N_f}}(\bar\psi_i\psi_i)^2.
  \label{eq:GN}
  \end{equation}
  For bare fermion mass $m=0$, there is a discrete chiral 
  symmetry
  \begin{equation}
  \psi\mapsto\gamma_5\psi\;\;\;;\;\;\;\bar\psi\mapsto-\bar\psi
 \gamma_5,
  \end{equation}
  which is spontaneously broken whenever a non-vanishing 
  condensate
  $\langle\bar\psi\psi\rangle$ is generated. To proceed, we 
  introduce a 
  bosonic scalar auxiliary field $\sigma$ and rewrite
  \begin{equation}
  {\cal L}_{GN}=\bar\psi_i(\partial{\!\!\! 
  /}\,+m+\sigma)\psi_i
  +{N_f\over{2g^2}}\sigma^2.
  \label{eq:GNaux}
  \end{equation}
  The original Lagrangian (\ref{eq:GN}) can be recovered by 
  Gaussian
  functional integration over $\sigma$. Chiral symmetry 
  breaking 
  for $m\to0$ is now
  signalled by a non-vanishing vacuum expectation value
  $\Sigma\equiv\langle\sigma\rangle$ for the scalar field: 
  it then follows
  from (\ref{eq:GNaux}) that the fermion gets a dynamically 
  generated mass
  $M\simeq\Sigma$. 
  
  We can calculate $\Sigma$ using an expansion in inverse
  powers of $N_f$, the number of flavors. This expansion 
  associates a 
  factor $N_f$ with each closed fermion loop, and in effect $1/\surd 
  N_f$ with each
  fermion-scalar interaction vertex. To leading order, in 
  the chiral limit
  $m\to0$, only the tadpole diagram shown in Fig.~1a 
  contributes to
  $\Sigma$, leading to the self-consistent {\sl Gap 
  Equation\/}:
  \begin{equation}
  {\Sigma\over g^2}=\int_p\mbox{tr}{1\over{ip{\!\!\!
  /}\,+\Sigma}},
  \label{eq:gap}
  \end{equation}
  or, with a simple UV cutoff $\Lambda$ on momentum (note 
  $2<d<4$)
  \begin{equation}
  {1\over g^2}={8\over{(4\pi)^{d\over2}(d-2)}}
  \left[{\Lambda^{d-2}\over{\Gamma({d\over2})}}-\Sigma^{d-2}
  \Gamma(2-\textstyle{d\over2})\right].
  \label{eq:gapsol}
  \end{equation}
  
  \begin{figure}
  \centerline{
  \psfig{figure=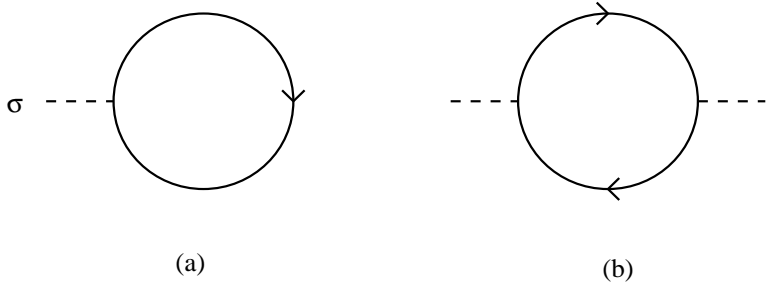,width=4.0in}
  }
  \caption{Leading order diagrams in the GN model}
  \end{figure}
  
  Equation (\ref{eq:gapsol}), relating a bare coupling 
  constant $g^2$ to
  both a UV scale $\Lambda$ and a physical scale $\Sigma$, 
  can be
  interpreted as a renormalisation condition. It turns out 
  to be the
  physical solution of (\ref{eq:gap}) for couplings larger 
  than the
  critical coupling $g_c^2$ given by
  \begin{equation}
  {1\over
  g_c^2}={{8\Lambda^{d-2}}\over{(4\pi)^{d\over2}(d-2)\Gamma(
  {d\over2})}},
  \end{equation}
  at which point chiral symmetry is spontaneously broken 
  (see Fig.~4).
  Only for $g^2\to g_c^2$, can the 
  ratio $\Lambda/\Sigma$ be made to diverge, implying a 
  continuum limit.
  Note that it is also possible to approach the continuum 
  limit from the
  symmetric phase~\cite{KikYam}.
  
  Remarkably, for $2<d<4$, to the same leading order in 
  $1/N_f$ all
  dependence on $\Lambda$ is absorbed in defining the value 
  of $g_c^2$.
  The only other Green function involving a fermion loop is 
  the scalar
  two-point function shown in Fig.~1b, but in the vicinity of the fixed 
  point, once
  (\ref{eq:gapsol}) is taken into account the scalar 
  propagator is finite
  and can be expressed in closed form~\cite{HKK}. 
Eg, for 
  $d=3$,
  \begin{equation}
  D_\sigma(k)={1\over N_f}{{2\pi\surd 
  k^2}\over{(k^2+4\Sigma^2)\tan^{-1}\left(
  {{\surd{k^2}}\over{2\Sigma}}\right)}}.
\label{eq:Dsigma}
  \end{equation} 
  In the large-$N_f$ limit $D_\sigma$ is essentially a 
  chain of fermion
  bubbles. It is worth examining its behaviour in two 
  limits. In the
  infra-red,
  \begin{equation}
  \displaystyle{\lim_{k\to0}}D_\sigma(k)\propto{1\over{k^2+4
  \Sigma^2}},
  \end{equation}
  and hence the $\sigma$ resembles a fundamental boson with 
  mass
  $2\Sigma$. This shows the scalar to be a weakly bound 
  fermion -
  anti-fermion composite. In the ultra-violet, on the other 
  hand, for
  general $d$ we have
  \begin{equation}
  \displaystyle{\lim_{k\to\infty}}D_\sigma(k)\propto{1\over{
  k^{d-2}}}.
  \end{equation}
  Thus the UV asymptotic behaviour is {\sl harder\/} than 
  that of a 
  fundamental scalar, but still {\sl softer\/} than the 
  $1/k^0$
  corresponding to a non-propagating auxiliary field. 
  The strong interaction between 
  fermion and anti-fermion is responsible for this 
  modification,
  which in turn makes diagrams corresponding to higher 
  order corrections,
  such as those of Fig.~2, less divergent than expected by 
  naive power
  counting.
  
  \begin{figure}
  \centerline{
  \psfig{figure=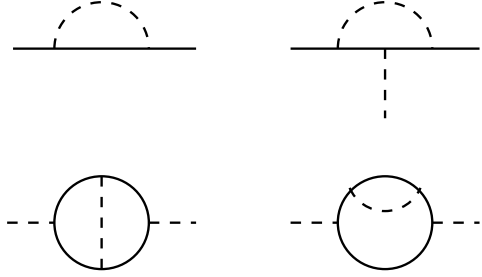,width=2.5in}
  }
  \caption{$O(1/N_f)$ radiative corrections in the GN 
  model}
  \end{figure}
  
  The result, known for a long time~\cite{WilGro},
  but discussed recently with renewed interest~\cite{RWP},
  is that the $1/N_f$ expansion about the fixed point 
  $g^2=g_c^2$ is
  exactly renormalisable for $2<d<4$.
  
  An important aspect of the model's renormalisability is 
  that it
  depends on a precise cancellation between logarithmic 
  divergences from
  different graphs~\cite{HKK}. This can be understood in 
  terms of the four-fermi
  scattering amplitude, shown schematically in Fig.~3: 
  there are three different types of logarithmic 
  divergence, each represented by a blob, but in the 
  massless limit only two tuneable parameters in the 
  renormalised Lagrangian, implying a non-trivial 
  consistency relation. 

\begin{figure}
  \centerline{
  \psfig{figure=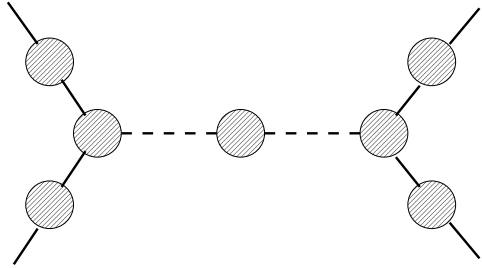,width=2.5in}
  }
  \caption{Renormalised four-fermi scattering amplitude}
  \end{figure}
  In turn this implies that in the deep UV
  limit the amplitude assumes a universal form
  \begin{equation}
  \displaystyle{\lim_{k\to\infty}}{\cal M}_{ff\to ff}=
  {A_d\over{N_f k^{d-2}}},
  \end{equation}
  with the numerical constant 
  $A_d=(4\pi)^{d\over2}/4\Gamma(2-{d\over2})
  B({d\over2},{d\over2}-1)$, and crucially {\sl no\/} 
  $1/N_f$
  corrections. How then do higher order corrections 
  manifest themselves?
  Fig.~4, showing a comparison between Monte Carlo data for 
  $\Sigma$ from
  a $12^3$ lattice with the predictions of the leading 
  order gap equation
  (\ref{eq:gap}), gives a clue. 
\begin{figure}
  \centerline{
  \psfig{figure=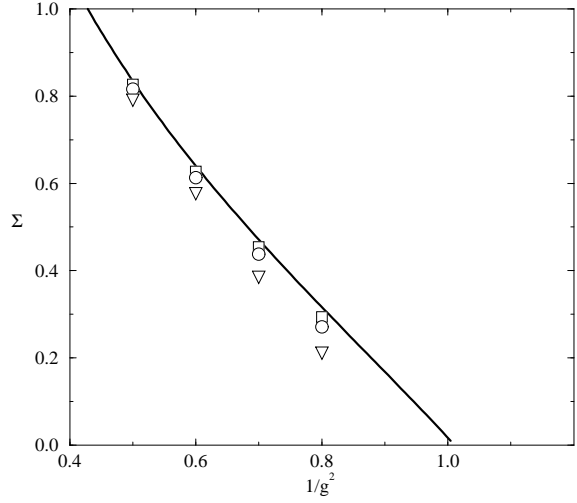,width=3.5in}
  }
  \caption{Plot of $\Sigma$ vs. $1/g^2$ for $N_f=24$ (squares),
12 (circles), and 6 (triangles). The line is the leading order
gap equation prediction.}
  \end{figure}
The corrections to $\Sigma$ 
  are clearly
  $O(1/N_f)$, suggesting that it is the behaviour of the 
  bulk
  ``thermodynamic'' observables near the phase transition, 
  in other words
  the IR physics, which most conveniently expresses the 
  higher order information. If we define
  a {\sl reduced coupling\/}
  \begin{equation}
  t={1\over g^2}-{1\over g_c^2},
  \end{equation}
  then the scaling of the order parameter $\Sigma$ near the 
  fixed point may
  be described by a set of {\sl critical exponents\/}.
  Explicitly~\cite{HKK}~\cite{He}:
  \begin{eqnarray}
  \Sigma\propto t^\beta &\mbox{with}& \beta={1\over{d-2}}+
  O\left({1\over
  N_f^2}\right);\\
  \Sigma\vert_{t=0}\propto m^{1\over\delta} &\mbox{with}&
  \delta=(d-1)\left[1+{C_d\over 
  N_f}\right];\label{eq:delta}\\
  \chi\equiv{{\partial\Sigma}\over{\partial 
  m}}\biggr\vert_{m=0}\propto\vert
  t\vert^{-\gamma} &\mbox{with}&
  \gamma=1+{{d-1}\over{d-2}}{C_d\over N_f};\\
  M\propto\vert t\vert^\nu &\mbox{with}& \nu={1\over{d-2}}
  \left[1+{{d-1}\over d}{C_d\over N_f}\right];\\
  D_\sigma(k)\vert_{t=0}\propto{1\over{k^{2-\eta}}} 
  &\mbox{with}&
  \eta=4-d-2{{d-1}\over d}{C_d\over N_f}.
  \end{eqnarray}
  The numerical constant 
  $C_d=1/B({d\over2},2-{d\over2})B({d\over2},{d\over2}-1)$.
  Several comments can be made:
  
  \begin{enumerate}
  
  \item[$\bullet$] The exponents obey certain consistency 
  checks known as
  {\sl scaling\/} and {\sl hyperscaling\/} relations, eg.
  \begin{equation}
  \gamma=\nu(2-\eta)\;\;\;;\;\;\;\beta={1\over2}\nu(d-2+\eta
  ).
  \end{equation}
  In statistical mechanics these are derived~\cite{Ma} 
  on the assumption that there
  is a single length scale characterising the important 
  physics, namely
  the correlation length $\xi$. With $\xi\sim{\Lambda\over 
  M}$, this is
  precisely the statement of renormalisability~\cite{HKK}.
  
  \item[$\bullet$] In the limit $d\to4$ the exponents 
  assume their
  mean field (Landau-Ginzburg) values, namely
  \begin{equation}
  \beta={1\over2}\;\;;\;\;\delta=3\;\;;\;\;\gamma=1\;\;;\;\;
  \nu={1\over2}
  \;\;;\;\;\eta=0.
  \label{eq:MFT}
  \end{equation}
  
  \item[$\bullet$] The exponent $\eta$ is related to the 
  anomalous 
  scaling dimension $\gamma_{\bar\psi\psi}$ of the 
  composite operator
  $\bar\psi\psi$~\cite{KikYam}:
  \begin{equation}
  \eta=d-2\gamma_{\bar\psi\psi}.
  \end{equation}
  This implies for the scaling dimension $[\bar\psi\psi]$ 
  the following:
  \begin{eqnarray}
  [\bar\psi\psi] &=& 
  [\bar\psi\psi]_0-\gamma_{\bar\psi\psi}\nonumber\\
  &=& d-1-(d-2+O(1/N_f))\nonumber\\
  &=& 1+O(1/N_f).
  \end{eqnarray}
  Hence the interaction term $(\bar\psi\psi)^2$ has scaling 
  dimension
  $\simeq2$, and is super-renormalisable, or in the 
  renormalisation
  group (RG) sense {\sl relevant\/} at the fixed 
  point~\cite{GKR}. Note 
  that the sign of the $O(1/N_f)$ correction is not crucial 
  to the
  argument~\cite{HKK}.
  
  \item[$\bullet$] With the addition of 
  $(\partial_\mu\sigma)^2$ and
  $\sigma^4$ terms, the model (\ref{eq:GNaux}) becomes a 
  Higgs-Yukawa
  (HY) theory, which is super-renormalisable for $d<4$. The 
  UV fixed point
  of the GN model can thus be viewed as an IR fixed point 
  of the HY
  model, at which the two extra operators become 
  irrelevant~\cite{Shizuya}.
  This recalls the Wilson-Fisher fixed point in 
  ferromagnetic models, and
  suggests an expansion in $\varepsilon=4-d$ as an 
  alternative approach to
  the study of the universality class~\cite{Zinn}. The 
  equivalence of the
  two models in $d=3$ has been tested 
  numerically~\cite{Focht}.
  
  \end{enumerate}
  
  The final issue we should address for the GN model is the 
  applicability
  of the $1/N_f$ expansion; in other words, for how small a 
  value of $N_f$
  may we trust its accuracy? Calculations of critical 
  exponents are now
  available to $O(1/N_f^2)$, and are compared to 
  predictions
  from the $\varepsilon$-expansion of the HY model to
  $O(\varepsilon^2)$, and the results of a $d=3$ Monte 
  Carlo 
  simulation for $N_f=2$ 
  (the smallest value of $N_f$ that is simulable 
  with a local lattice action using the staggered fermion 
  formulation, which retains a chiral 
  symmetry~\cite{Burden}\footnote{
  This formulation has a parity-invariant mass term, 
  consistent with our use of four-component 
  spinors~\cite{Luscher}.}) 
  in Tab.~\ref{tab:GNZ2}.
  \begin{table}[t]
  \setlength{\tabcolsep}{1.5pc}
  \caption{Critical exponents for the $d=3$ $N_f=2$ 
  Gross-Neveu model}
  \label{tab:GNZ2}
  \vspace{0.4cm}
  \begin{center}
  \begin{tabular}{|clll|}
  \hline
  &   
  large-$N_f$~\cite{jag1}&$\varepsilon$-expansion~\cite{Kark}
  & Monte Carlo~\cite{Kark}  \\
  \hline
  $\nu$              &  0.903    &  0.9480  &  1.00(4)     
  \\
  $\gamma/\nu$   &  1.2559    &  1.237  &  1.246(8)   \\
  \hline
  \end{tabular}
  \end{center}
  \end{table}
  The agreement is satisfactory, though it should be noted 
  that the
  $1/N_f$ expansion for $\nu$, and the 
  $\varepsilon$-expansion for
  $\gamma/\nu$ appear slowly if at all convergent, and the 
  values quoted
  in the table have been obtained using resummation
  techniques~\cite{Kark}.
  
  A similar comparison between the large-$N_f$ expansion
  to $O(1/N_f^2)$ and Monte Carlo results has also
  been made for the GN model with continuous U(1) chiral 
  symmetry,
  for $N_f=4$, with the results shown in 
  Tab.~\ref{tab:GNU1}.
  \begin{table}[t]
  \setlength{\tabcolsep}{1.5pc}
  \caption{Critical exponents for the $d=3$ $N_f=4$ 
  Gross-Neveu model
  with U(1) chiral symmetry}
  \label{tab:GNU1}
  \vspace{0.4cm}
  \begin{center}
  \begin{tabular}{|cll|}
  \hline
  &   large-$N_f$~\cite{jag2}   &   Monte 
  Carlo~\cite{Focht}   \\
  \hline
  $\nu$              &  1.0(1) &  1.02(8)    \\
  $\gamma/\nu$   &  1.055(13)  &  1.19(13)    \\
  $\beta/\nu$   &  0.973(6)  &  0.89(10)    \\
  \hline
  \end{tabular}
  \end{center}
  \end{table}
  In this case no resummation is attempted on the $1/N_f$ 
  series: instead
  the quoted error reflects the size of the last term.
  
  We conclude that the physical picture of dynamical 
  symmetry breaking in the GN model given by the 
  large-$N_f$ expansion is
  borne out by explicit simulations in $d=3$, and is 
  qualitatively
  correct. Quantitative accuracy is limited by the 
  apparently rather slow
  convergence of certain series, such as that for the 
  exponent $\nu$.
  
  \section{The Thirring Model for $d=3$}
  
  The next model to consider, which will turn out to be 
  more interesting,
  is the Thirring model, originally studied as a soluble model
for $d=2$ and $N_f=1$~\cite{Thirring}, but here generalised.
It differs from the GN model in 
  that the contact
  interaction is between conserved currents:
  \begin{eqnarray}
  {\cal L}_{Thir} &=& 
  \bar\psi_i(\partial{\!\!\! /}\,+m)\psi_i
  +{g^2\over{2N_f}}(\bar\psi_i\gamma_\mu\psi_i)^2\nonumber\\
  &=&
  \bar\psi_i(\partial{\!\!\! /}\,+iA{\!\!\! /}\,+m)\psi_i+
  {N_f\over{2g^2}}A_\mu^2,
  \label{eq:thirring}
  \end{eqnarray}
  where in the second line we have introduced an auxiliary 
  vector field
  $A_\mu$. The form of the first term of 
  (\ref{eq:thirring}) suggests that
  of an abelian gauge theory, except that the term 
  quadratic in $A_\mu$
  violates the gauge symmetry. In the chiral limit the Lagrangian 
  (\ref{eq:thirring}) has a continuous U(1) chiral 
  symmetry:
  \begin{equation}
  \psi\mapsto\exp(i\alpha\gamma_5)\psi\;\;\;;\;\;\;
  \bar\psi\mapsto\bar\psi\exp(i\alpha\gamma_5).
  \label{eq:chirsym}
  \end{equation}
  It is possible to develop the $1/N_f$
  expansion just as we did for the GN model, but this time 
  in the chiral
  limit $\langle\bar\psi\psi\rangle$ remains zero to all 
  orders,
  essentially because the trace over an odd number of 
  $\gamma$-matrices
  vanishes. Consequently there is no fixed-point condition 
  analogous 
  to (\ref{eq:gapsol}); however,
  provided a 
  regularisation is
  specified which preserves current conservation
  $\partial_\mu(\bar\psi\gamma_\mu\psi)=0$, the auxiliary
propagator $D_{\mu\nu}(k)$ is actually finite at leading order. For $d=3$ we 
  have, explicitly
  \begin{equation}
  {N_f\over g^2}D_{\mu\nu}(k)=\left(\delta_{\mu\nu}-{{k_\mu 
  k_\nu}\over{k^2}}\right)
  {1\over\displaystyle{1+{g^2\over{2\pi}}\left(m+{{k^2-4m^
  2}\over{2\surd
  k^2}}\tan^{-1}\left({{\surd 
  k^2}\over{2m}}\right)\right)}}
  +{{k_\mu k_\nu}\over k^2}.
  \label{eq:Dmunu}
  \end{equation}
  
  The longitudinal piece has no physical content, since it 
  vanishes on
  being applied to a conserved current. As before, we can 
  examine
  (\ref{eq:Dmunu}) in two opposite limits. In the UV, we 
  find
  \begin{equation}
  \displaystyle\lim_{k\to\infty}D_{\mu\nu}(k)\propto
  \left(\delta_{\mu\nu}-{{k_\mu 
  k_\nu}\over{k^2}}\right){1\over
  {k^{d-2}}}.
  \end{equation}
  Once again, the modified asymptotic behaviour means that 
  the divergence
  structure of higher order corrections is better behaved 
  than expected:
  the Thirring model is thus exactly renormalisable for 
  $2<d<4$~\cite{Paris}~\cite{Gomes}~\cite{Hands}\footnote
  {Strictly
  renormalisablity beyond leading order has only been demonstrated for the 
  massless
  theory~\cite{Hands}.}.
  In the IR limit, the position of the pole in the 
  transverse piece
  of $D_{\mu\nu}$ yields the mass $M_V$ of the propagating 
  boson in the
  vector channel. For $d=3$ we have~\cite{Hands}
  \begin{equation}
  M_V=\cases 
  {2m\left(1-2\exp\left(-{{2\pi}\over{mg^2}}\right)\right)
  &$mg^2\to0$;\cr
  m\sqrt{{6\over{mg^2}}}&$mg^2\to\infty$.\cr}
  \end{equation}
  The result depends in a non-trivial way on the 
  dimensionless combination
  $mg^2$. In the weak coupling limit $mg^2\to0$ the boson 
  is a
  weakly bound fermion -- anti-fermion state as in the GN 
  model; however
  in the strong coupling limit its mass vanishes, yielding 
  a model in
  which a massless vector mediates interactions between 
  conserved currents.
  This is strongly suggestive that in this limit the 
  Thirring model is 
  identical to $\mbox{QED}_3$. Indeed, in general we expect 
  the UV limit of the
  Thirring model to coinicide with the IR limit of 
  $\mbox{QED}_3$, 
  since in these limits the $1/N_f$ expansions of the two 
  models 
  coincide~\cite{Hands}~\cite{Espriu}.
  
  We can summarise the distinct features of the Thirring 
  model in $2<d<4$,
  as compared to the generic GN behaviour described in the 
  previous
  section, with a number of equivalent statements:
  
  \begin{enumerate}
  
  \item[$\bullet$] There is no fixed-point condition.
  
  \item[$\bullet$] The bare parameter $g^2$ is not 
  renormalised.
  
  \item[$\bullet$] The continuum theory contains a free 
  dimensionless
  parameter $mg^2$.
  
  \item[$\bullet$] The interaction 
  $(\bar\psi\gamma_\mu\psi)^2$ is
  {\sl marginal\/} in the RG sense.
  
  \end{enumerate}
  
  Is this the whole story? All of these statements 
  are made
  plausible in the context of the large-$N_f$ limit. We 
  will now argue
  that the behaviour of the Thirring model is radically 
  different for
  small $N_f$, if methods which are non-perturbative in 
  $N_f$ are used.
  
  The Schwinger-Dyson (SD) 
  approach~\cite{Gomes}~\cite{Itoh}~\cite{Hong}
  is to solve for the full inverse fermion propagator
  $S_F^{-1}=A(k)(ik{\!\!\! /}\,+M(k))$ self-consistently 
  using integral equations, shown diagramatically in Fig.~5. 
  
  \begin{figure}
  \centerline{
  \psfig{figure=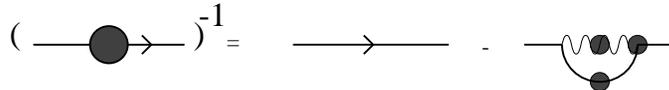,width=3.5in}
  }
  \caption{Schwinger-Dyson equation for the full fermion propagator}
  \end{figure}
  
Any non-trivial solution $M(k)\not=0$ in the chiral 
  limit signals dynamical mass generation. One 
  analysis~\cite{Itoh}
  exploits a hidden local symmetry of (\ref{eq:thirring}) 
  to fix a non-local ``gauge'' in which the wavefunction 
  renormalisation 
  $A(k)\equiv1$. To proceed, the SD equations must be 
  truncated;
  the full vector propagator is set to its large-$N_f$ form
  (\ref{eq:Dmunu}), and the full vertex $\Gamma_\mu(p,q)$ 
  to the 
  bare vertex $i\gamma_\mu$ -- this second step is known as 
  the ladder approximation, since in effect it ignores 
  diagrams with 
  crossed vector lines. With these approximations the SD 
  equation may be solved in closed form in the strong 
  coupling limit $g^2\to\infty$.
  
  The result is that a non-trivial solution is found for 
  small values
  of $N_f$; explicitly, for $d=3$
  \begin{equation}
  {M\over\Lambda}\propto\exp\left(-{{2\pi}\over\sqrt{
  {N_{fc}\over N_f}-1}}\right)\;\;\Rightarrow
  \langle\bar\psi\psi\rangle\vert_{m=0}\propto\Lambda^2\exp
  \left(
  -{{3\pi}\over\sqrt{\textstyle{N_{fc}\over 
  N_f}-1}}\right),
  \label{eq:SDsol}
  \end{equation}
  with $\Lambda$ a UV cutoff. The solution exists for
  \begin{equation}
  N_f<N_{fc}={{128}\over{3\pi^2}}\simeq4.3.
  \end{equation}
  If $N_f$ were a continuous parameter, then (\ref{eq:SDsol}) 
  would
  predict a UV fixed point as $N_f\to N_{fc}$. The 
  essentially singular behaviour is similar to that of 
  quenched 
  $\mbox{QED}_4$~\cite{Mir}; this type of transition has 
  been given the label ``conformal phase 
  transition''~\cite{MirYam}. How are we
  to interpret it in this case? It has been 
  argued~\cite{Itoh}~\cite{Kondo} that for $g^2<\infty$ the 
  critical curve $N_{fc}(g)$ is smooth. It is tempting to 
  propose the phase 
  diagram shown in Fig.~6, implying a sequence of fixed 
  points 
  $g^2_c(N_f)$ for integer values of $N_f<N_{fc}$. 
  The Thirring interaction 
  $(\bar\psi\gamma_\mu\psi)^2$ has become relevant.
  \begin{figure}
  \centerline{
  \psfig{figure=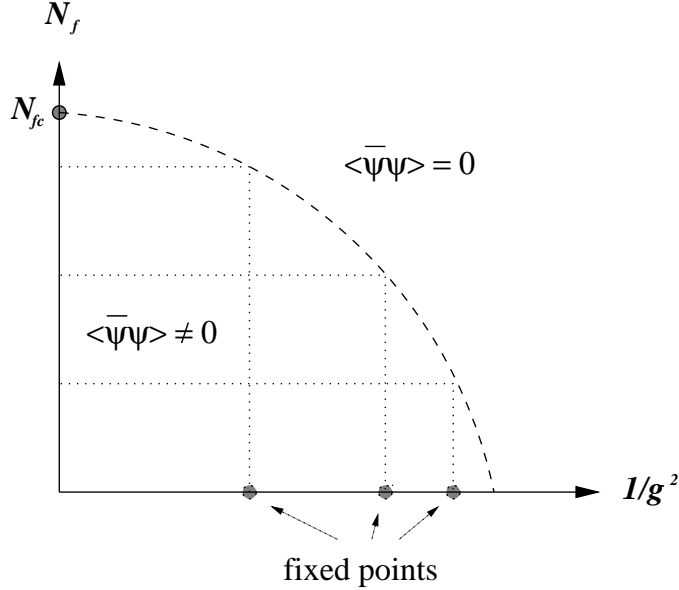,width=3.5in}
  }
  \caption{Proposed UV fixed points for the small-$N_f$ Thirring model}
  \end{figure}
  We should be cautious, however; a different sequence of
  truncations predicts $N_{fc}=\infty$~\cite{Hong}. Indeed,
  the situation is reminiscent of $\mbox{QED}_3$, where
chiral symmetry breaking is predicted for $N_f<N_{fc}$; the 
  predicted value of $N_{fc}$ being highly sensitive to the 
  assumptions
  used in solving the SD equation~\cite{QED3}~\cite{Mavro}. 
  Once
  again, however, the same SD equation appears to describe 
  both models, but in opposite momentum regimes, suggesting 
  that their 
  equivalence may hold beyond the large-$N_f$ 
  approximation.
  
  The UV behaviour of the Thirring model, if non-trivial, 
  is genuinely non-perturbative, since there is no small 
  dimensionless parameter in play at the putative fixed 
  points. In this sense the
  Thirring model resembles the strong-coupling behaviour of 
  $\mbox{QED}_4$~\cite{QED4}~\cite{KKW}, and hence is a suitable case 
  for lattice Monte Carlo treatment. Indeed, to the best of 
  my knowledge the
  three dimensional Thirring model may be the simplest 
  fermionic model requiring a numerical solution, and thus 
  deserves study by lattice theorists interested in 
  simulating dynamical lattice fermions.
  
  Lattice studies of the Thirring model have been made by 
  two groups;
  I will briefly summarise our results~\cite{us}: 
  details of the other group's approach are given by the 
  next speaker.~\cite{Kim}
  The lattice formulations we have used differ in detail. I 
  will suppress technical details, 
  though it is still not clear how important they might turn 
  out to be; one might hope that the number of distinct 
  universality classes for fixed
  $N_f$ consistent with the breaking of the symmetry 
  (\ref{eq:chirsym}) is small. We have studied lattice 
  models 
  corresponding to $N_f=2$, 4 and 6 on volumes $8^3$, 
  $12^3$
  and $16^3$, with bare fermion masses 
  $ma=0.05,\ldots,0.01$
  ($a\sim\Lambda^{-1}$ is the lattice spacing).
  
  First let's discuss the evidence for spontaneous chiral 
  symmetry 
  breaking. We have measured 
  $\Sigma=\langle\bar\psi\psi\rangle$ for various $m$ and 
  $g^2$,
  and fitted our data to the equation of state (EOS)
  \begin{equation}
  m=A\left({1\over g^2}-{1\over g_c^2}\right)\Sigma
  +B\Sigma^\delta.
  \label{eq:eos}
  \end{equation}
  The exponent $\delta$ is identical to that of 
  (\ref{eq:delta}) --
  the form (\ref{eq:eos}) makes the implicit assumption 
  that
  $\delta-1/\beta=1$, which is consistent with the ladder 
  approximation~\cite{us}. To explore (\ref{eq:eos}), it is
  convenient to display the data in the form of a Fisher 
  plot, ie.,
  $\Sigma^2$ versus $m/\Sigma$. For the mean field 
  exponents
  (\ref{eq:MFT}), this plot yields curves of constant $g^2$ 
  as evenly
  spaced parallel straight lines, intercepting the vertical 
  axis 
  for $g^2>g_c^2$, the horizontal axis for $g^2<g_c^2$, and 
  passing
  through the origin at the critical coupling. Any 
  deviation signals
  a departure from the predictions of mean field theory. In 
  Figs.~7,
8, and 9 we show Fisher plots respectively for 
  $N_f=2$, 4 and 6.
\begin{figure}
\epsfaxhax{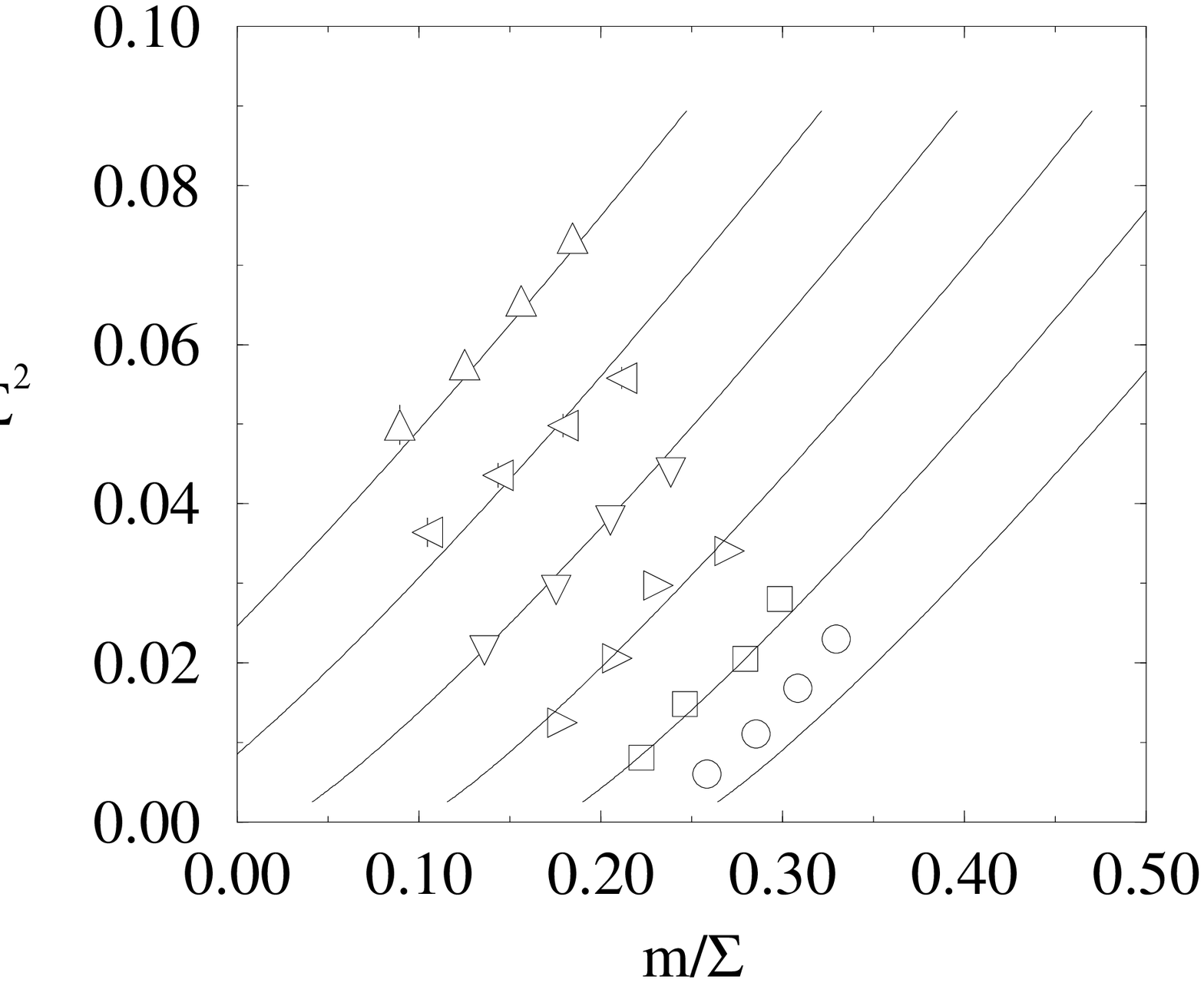}{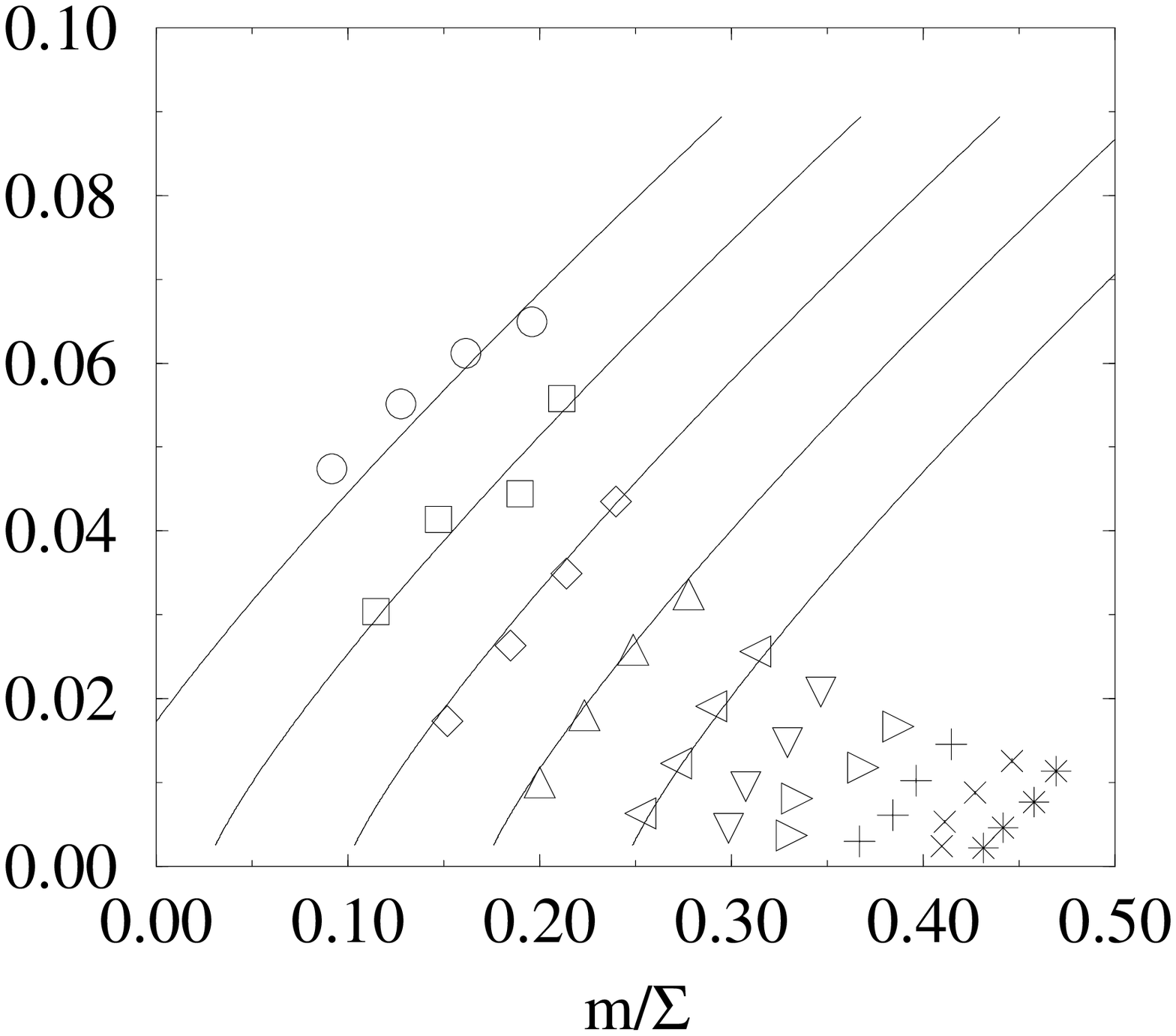}
\caption{\hskip 4.1cm Figure 8:}
\setcounter{figure}{8}
  \centerline{
  \psfig{figure=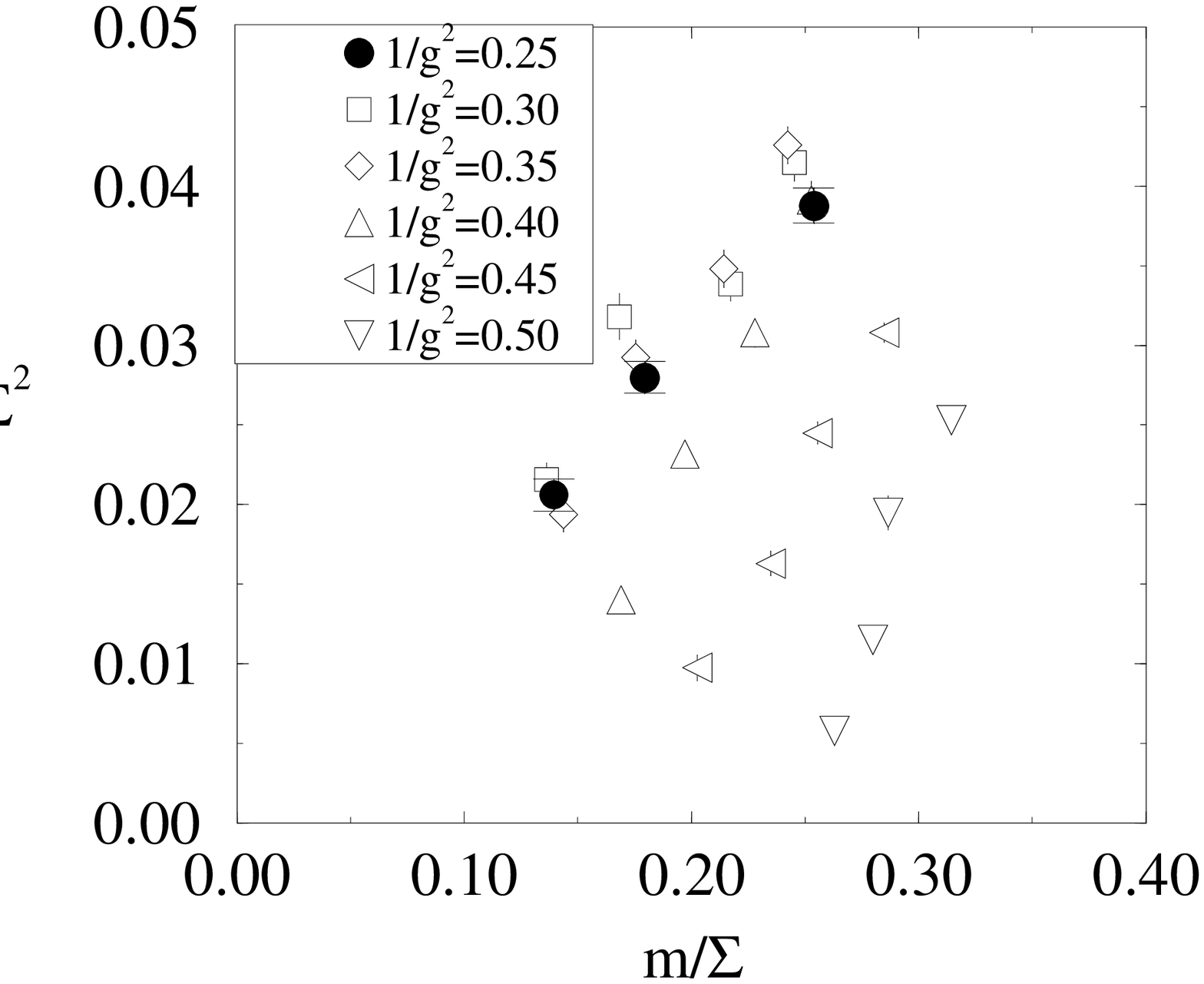,width=2.2in}
  }
\caption{}
  \end{figure}
The main features to note are that the $N_f=2,4$ data 
  show 
  different curvature, and that the $N_f=6$ data appear to 
  be
  accumulating in the sub-critical region. The results of a 
  fit to
  (\ref{eq:eos}), which include a finite volume scaling 
  analysis 
  for $N_f=2$, are summarised in Tab.~\ref{tab:fit}.
   
  \begin{table}[t]
  \setlength{\tabcolsep}{1.5pc}
  \caption{Fits to the Equation of State}
  \label{tab:fit}
  \vspace{0.4cm}
  \begin{center}
  \begin{tabular}{|cccc|}
  \hline
             &   $N_f=2$   &   $N_f=4$   & $N_f=6$  \\
  \hline
  $1/g_c^2$  &  1.92(2)    &  0.66(1) & no fit    \\
  $\delta$   &  2.75(9)    &  3.43(9)  & found  \\
  \hline
  \end{tabular}
  \end{center}
  \end{table}
  
  We conclude that the $N_f=2$ and $N_f=4$ models do have
  fixed points, and that even if we relax the assumptions
implied in (\ref{eq:eos}) they fall in distinct universality classes; 
  the data collected so far support $4<N_{fc}<6$,
  
Next consider susceptibilities, which are equivalent to spatially
integrated two-point functions. By analogy with ferromagnetic models
we define a ``longitudinal'' (ie. scalar) susceptibility $\chi_l$
and a ``transverse'' (ie. pseudoscalar) susceptibility $\chi_t$
as follows:
\begin{equation}
\chi_l=\sum_x\langle\bar\psi\psi(0)\bar\psi\psi(x)\rangle\;\;\;;\;\;\;
\chi_t=\sum_x\langle\bar\psi\gamma_5\psi(0)\bar\psi\gamma_5\psi(x)\rangle.
\end{equation}
The longitudinal susceptibility has contributions from diagrams with
both connected ($\sqcap\!\!\!\!\sqcup$) 
and disconnected ($\circ$) fermion lines (ie. flavor non-singlet
and singlet channels respectively), plotted in Fig.~10. 
\begin{figure*}
\epsfaxhax{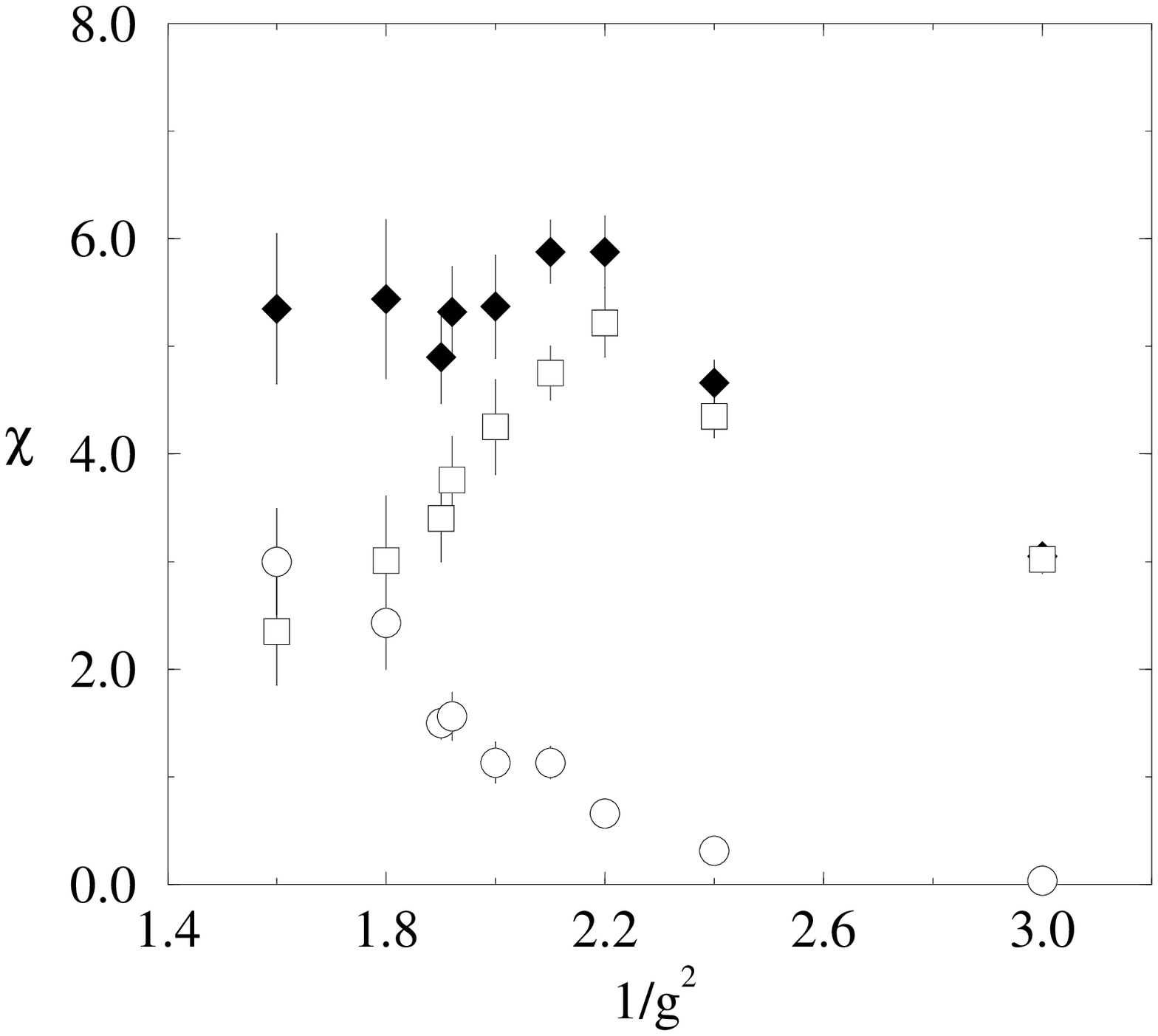}{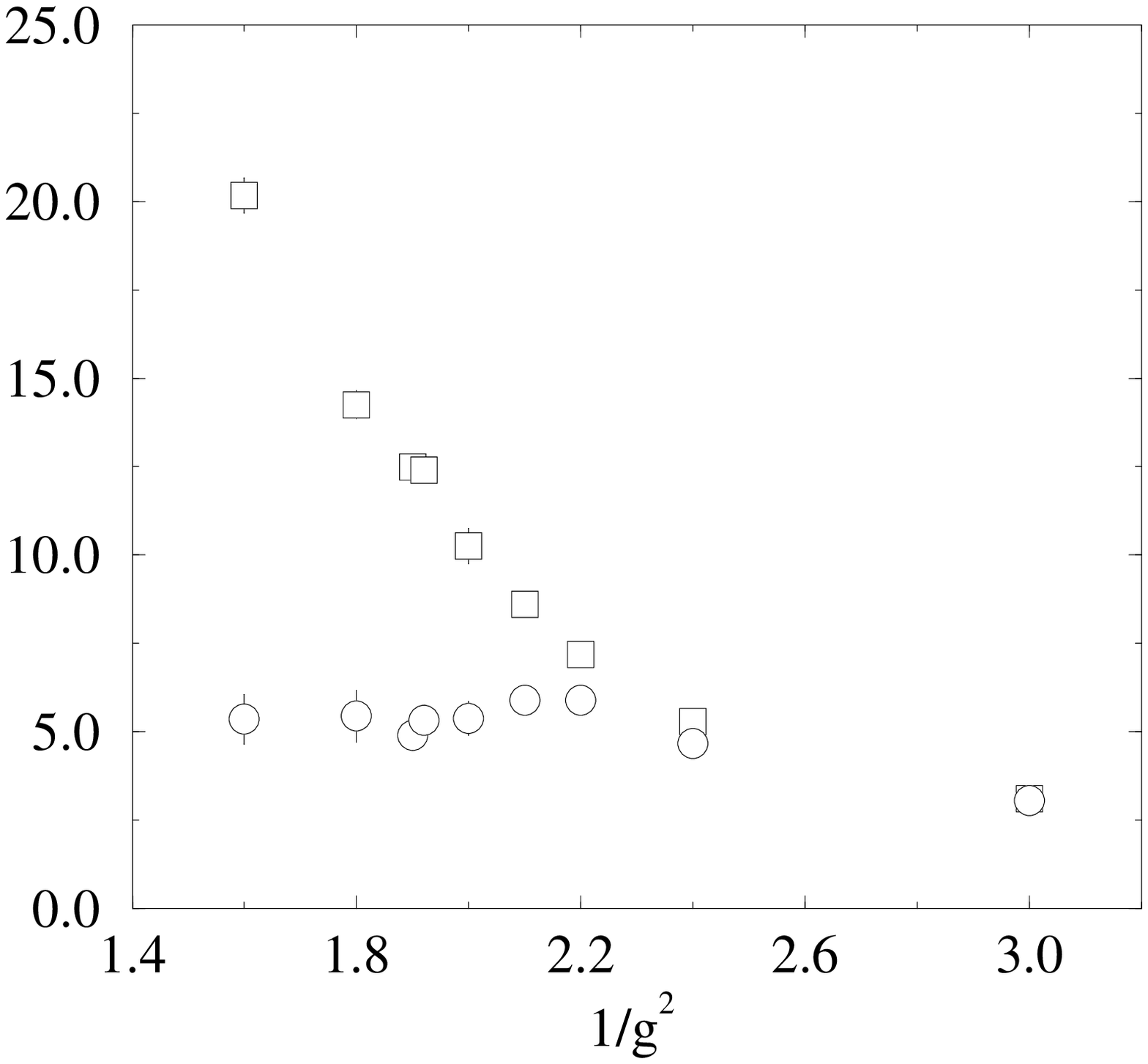}
\caption{\hskip 4.1cm Figure 11:}
\end{figure*}
The
relative importance of the disconnected piece, which is
entirely due to sea quark loops, rises significantly
in the broken symmetry phase. Fig.~11 shows that in the symmetric phase
$\chi_l$ ($\circ$) and $\chi_t$ ($\sqcap\!\!\!\!\sqcup$)
are approximately equal, as required by chiral
symmetry, whereas in the broken phase $\chi_t\gg\chi_l$, since
$\chi_t\propto M_\pi^{-2}$, and we expect the pion to be a Goldstone
boson in the chiral limit. The ratio $\chi_l/\chi_t=R(m,g^2)$ is 
another interesting quantity to examine. In general, for fixed $g^2$, 
$R$ varies strongly as a function of $m$ as $m\to0$, tending in the
chiral limit to zero in the broken phase and to a constant $\sim O(1)$ in
the symmetric phase. Exactly at the critical coupling, however, it
follows from the EOS (\ref{eq:eos}) and the chiral Ward identity
that~\cite{KKW}
\begin{equation}
R\vert_{g^2=g_c^2}={1\over\delta},
\end{equation}
independently of $m$. Thus numerical measurements of $R$ give an
independent estimate of the exponent $\delta$. In Fig.~12 we show data
for $N_f=2$ in the vicinity of the critical coupling, 
for $1/g^2=1.9$ ($\circ$), 1.92
($\sqcap\!\!\!\!\sqcup$), and 2.0 ($\diamond$),
together with 
the value of ${1\over\delta}$ from Tab.~\ref{tab:fit}. 
\begin{figure*}
\setcounter{figure}{11}
\epsfaxhax{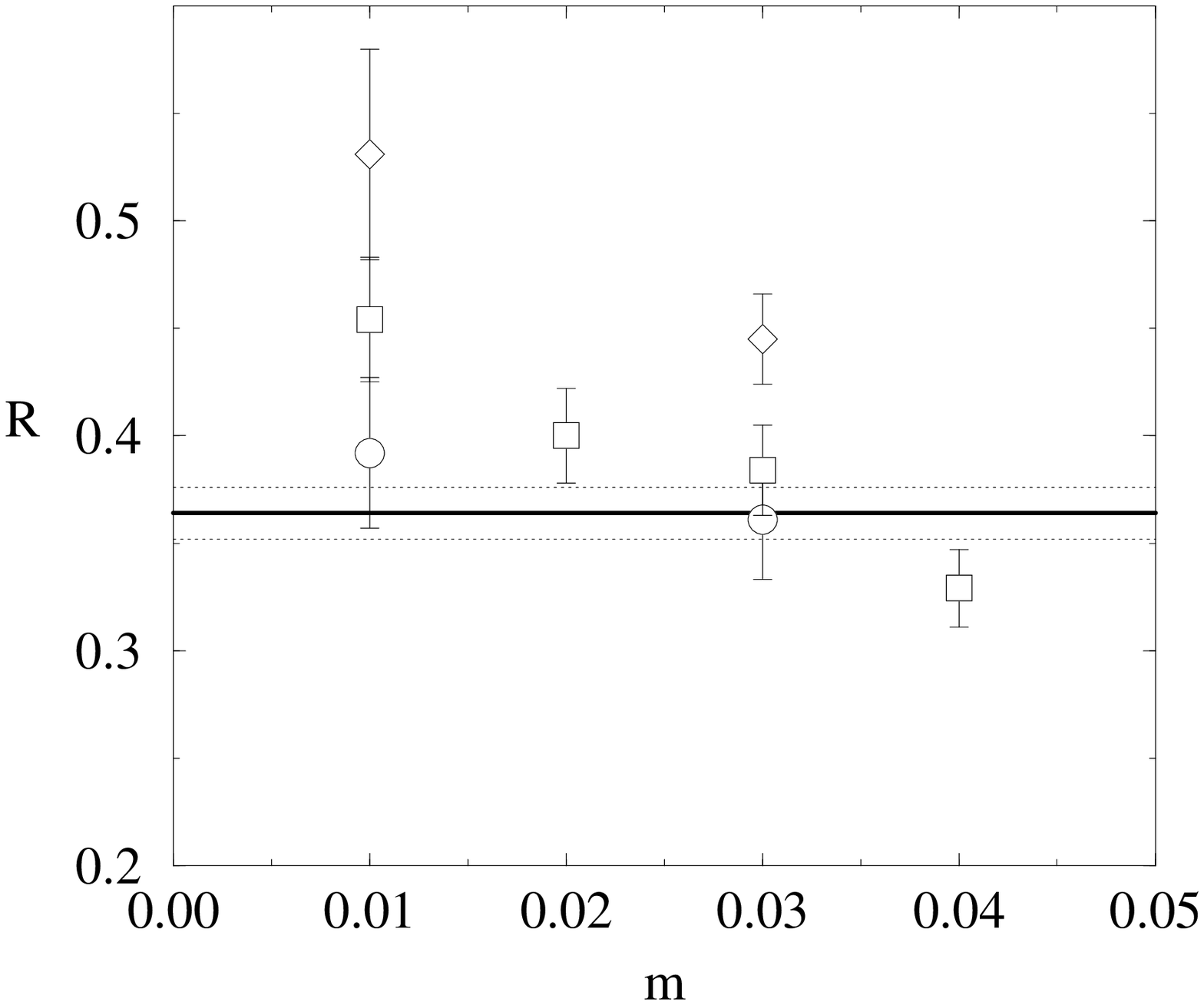}{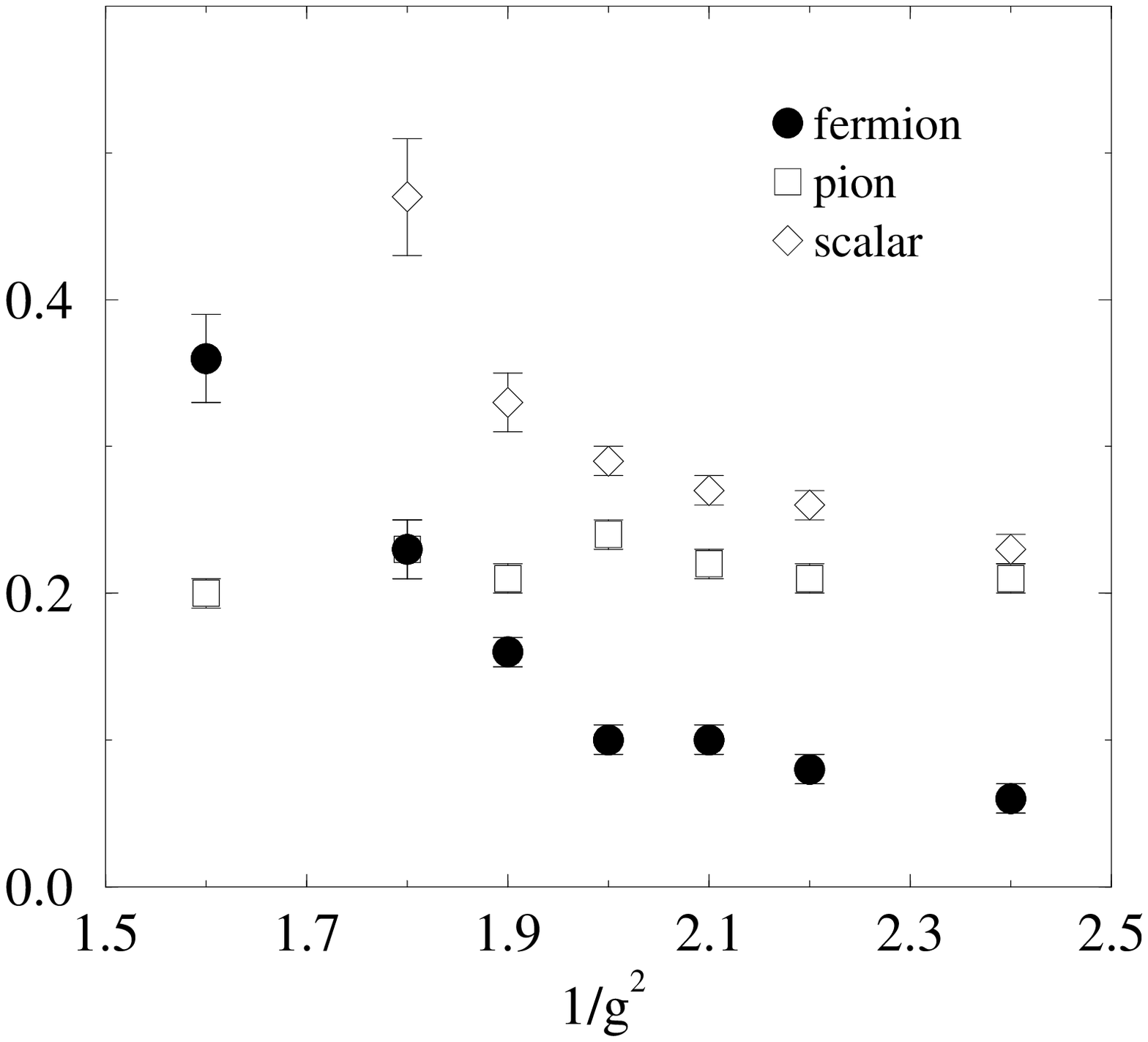}
\caption{\hskip 4.1cm Figure 13:}
\end{figure*}
The data suggest
that $R$ may become independent of $m$ at criticality once finite volume
effects are taken into account~\cite{us}. Whilst this result is
preliminary, it is interesting because measuring $R$ may be the most
promising method of testing the equivalence of the Thirring model
with $\mbox{QED}_3$ in future simulations.

Finally I present results on the mass spectrum for $N_f=2$,
both for the physical
fermion $f$ and for scalar and pseudoscalar $f\bar f$ bound states. 
A $16^3$ lattice is really too small in the timelike direction for any
great precision; however Fig.~13 demonstrates that the level ordering
of the states varies sharply across the transition. 
There is no evidence
for the absence of light composite states in the symmetric phase, as
predicted for the case of a conformal phase transition~\cite{MirYam}.
One further result is that in the vector channels in the
symmetric phase there are light states
in both ``vector'' ($\bar\psi\gamma_\mu\psi$) and ``axial vector''
($\bar\psi\gamma_\mu\gamma_5\psi$) channels: the latter is not present
in the large-$N_f$ approximation~\cite{us}.

So, the Thirring model is a good theoretical laboratory for
non-perturbative dynamical symmetry breaking. In future studies it
will be interesting to refine the spectroscopy, perhaps to explore the
effects of a strongly-interacting continuum above the $f\bar f$
threshold. More effort is still needed to pin down the value of
$N_{fc}$, and to see if the Thirring model and $\mbox{QED}_3$ 
are equivalent descriptions of the same critical dynamics. 

\section{The Nambu -- Jona-Lasinio Model for $d=4$}

After all these nice results, it is natural to demand what happens
in the physical dimension $d=4$. Once again we revert to the 
GN model in the large-$N_f$
expansion. As $d\to4$, the expansion loses the property of
renormalisability, essentially because the ``vacuum polarisation'' 
contribution to the auxiliary two-point function gets a subleading
$\ln\Lambda$ divergence. This must be compensated by the introduction of
a kinetic term $(\partial_\mu\sigma)^2$ in the renormalised Lagrangian.
If instead we choose to work with a finite cutoff $\Lambda$, then it is
possible to make do without introducing the scalar kinetic term, at the
cost of a residual logarithmic dependence on the UV scale. A moderate
separation of UV and IR scales is possible -- the model is thus an
effective theory, making continuum-like predictions but with
$O(\ln\Lambda)$ scaling violations. The UV scale can be interpreted
as that at which new physical information, in the form of a more
fundamental theory, must be supplied. The technical name for this 
scenario is {\sl triviality\/}~\footnote{Actually the model remains
trivial even once promoted to a Higgs-Yukawa theory with the addition 
of $(\partial_\mu\sigma)^2$ and $\sigma^4$ operators.} -- in the 
continuum limit $\Lambda\to\infty$ all interaction strengths vanish, 
as we shall see below. In RG language, for $d=4$ it is now the Gaussian fixed 
point which dominates the low-energy physics, and different models
differ only by logarithmic corrections from free field theory.

The model we will consider in detail is the Nambu -- Jona-Lasinio (NJL)
model~\cite{Nambu}:
\begin{equation}
{\cal L}_{NJL}=\bar\psi_{ip}(\partial{\!\!\! /}\,+m)\psi_{ip}
-{g^2\over{2N_f}}\left[(\bar\psi_{ip}\psi_{ip})^2
-(\bar\psi_{ip}\gamma_5\vec\tau_{pq}\psi_{iq})^2\right],
\end{equation}
where $p,q$ run over an internal isospace. In the chiral limit
${\cal L}_{NJL}$ has an $\mbox{SU(2)}_L\otimes\mbox{SU(2)}_R$ axial
symmetry, which at strong coupling 
is dynamically broken to $\mbox{SU(2)}_V$ by the
generation of a condensate $\langle\bar\psi\psi\rangle$.
This is similar to the pattern of symmetry breaking in the Higgs
sector of the Standard Model, and has led to suggestions that the 
physical Higgs is a $f\bar f$ composite~\cite{topmode}.
The field theory conventionally used to describe the Higgs sector
is, of course, the O(4) linear sigma model, describing elementary scalars
with a four-point interaction. Both theories are trivial, and once the
extra operators corresponding to scaling violations are taken into
account, then each has equivalent predictive power for the Standard
Model~\cite{Has}. That said, it is interesting to ask whether there
is a distinction between theories such as the sigma model in which
triviality is manifested through elementary scalars, and theories
like the NJL model in which fermions are elementary but scalars are
composite.

One way of understanding the distinction is via the wavefunction
renormalisation constant $Z$, defined as the coefficient of $1/k^2$ in
the scalar propagator in the IR limit $k^2\to0$. For the sigma model,
$Z$ is perturbatively close to one, whereas for the NJL model in the
large-$N_f$ expansion, for $d<4$ $Z\propto\Sigma^{4-d}$ (eg.
(\ref{eq:Dsigma})), and for $d=4$ $Z\propto1/\vert\ln\Sigma\vert$.
In either case $Z$ vanishes in the continuum limit $\Sigma\to0$,
with $\Sigma$ measured in cutoff units: this is the {\sl compositeness
condition\/}~\cite{Shizuya}. Another difference lies in the relative
ordering of scales. The interaction strength in the scalar sector
is expressed in terms of the pion decay constant $f_\pi$, which must
be finite for interactions to occur. For the sigma model,
\begin{equation}
{{f_\pi}\over\Sigma}\sim O(1),
\end{equation}
however the ratio between the order parameter $\Sigma$ and the physical
Higgs mass $M_\sigma$ is
\begin{equation}
{\Sigma\over M_\sigma}\sim{1\over\sqrt{g_R}}\sim\vert\ln\Sigma
\vert^{1\over2},
\end{equation}
where $g_R$ is the renormalised $\phi^4$ coupling constant, which 
vanishes logarithmically in the continuum limit. For the NJL model, 
on the other hand, to leading order in $1/N_f$ we have
~\cite{KK}~\cite{HK}
\begin{equation}
{f_\pi\over\Sigma}\sim\vert\ln\Sigma\vert^{1\over2}\;\;\;;\;\;\;
{\Sigma\over M_\sigma}\sim O(1).
\end{equation}
All these predictions follow from the PCAC relation $Zf_\pi^2=\Sigma^2$.
In both cases, triviality is realised  by the divergence of $f_\pi$
in units of the physical scale $M_\sigma$ in the continuum limit:
\begin{equation}
{f_\pi\over M_\sigma}\sim\vert\ln\Sigma\vert^{1\over2}.
\end{equation}

The different triviality scenarios also predict different forms for 
the equation of state. Recall that in $d=4$ the critical exponents 
take mean field values (\ref{eq:MFT}). Due to triviality the terms
in the EOS (\ref{eq:eos}) are modified by logarithmic corrections, to
give the form
\begin{equation}
m=A\left({1\over g^2}-{1\over
g_c^2}\right){\Sigma\over{\vert\ln\Sigma\vert^{q_1}}}+
B{\Sigma^3\over{\vert\ln\Sigma\vert^{q_2}}}.
\label{eq:eosmft}
\end{equation}
In the sigma model, the powers $q_1$ and $q_2$ can be found
analytically by using the RG to evolve the coupling from a generic
value into the perturbative regime~\cite{Zinn2}: this works because
even the perturbative EOS predicts symmetry breaking. For
the NJL model, by contrast, symmetry breaking only occurs for
large coupling strengths, where the only guide is the large-$N_f$
approach~\cite{KK}. The predictions are shown in Tab.~\ref{tab:mft+log} 
(the two RG values quoted for $q_1$ correspond
to U(1) and SU(2) symmetries respectively).
   
  \begin{table}[t]
  \setlength{\tabcolsep}{1.5pc}
  \caption{The correction exponents $q$ for $d=4$}
  \label{tab:mft+log}
  \vspace{0.4cm}
  \begin{center}
  \begin{tabular}{|lcccc|}
  \hline
    &   sigma & NJL   & `Thirring'~\cite{AliK} &  `GN'~\cite{HK} \\
  \hline
  $q_1$  &  0.4, 0.5    &  0 & 0.36(11) & 0.016(11)   \\
  $q_2$  &  1    &  $\!\!\!\!\!-1$  &  $\,\,\,0.485(17)$ & $\!\!\!\!\!
-0.553(18)$ \\
  \hline
  \end{tabular}
  \end{center}
  \end{table}

Corrections to the mean field EOS have been analysed in several 
recent studies~\cite{KKK}~\cite{AliK}~\cite{HK}, which use two different
lattice implementations of the four-fermi interaction. The first, 
which in $d=3$ we have used to define the lattice GN model, formulates
the auxiliary fields $\sigma$ on the sites of the dual
lattice~\cite{Cohen}, so that each $\bar\psi\psi$ pair interacts with
all other sites sharing a hypercube. This has been used in a study 
of the $N_f=2$ NJL model with $\mbox{SU(2)}_L\otimes\mbox{SU(2)}_R$
symmetry~\cite{HK}. The second, which in $d=3$ defines the lattice
Thirring model, formulates the auxiliaries on the lattice links
~\cite{BKP}; the $N_f=4$ model with U(1) chiral symmetry was then
studied~\cite{AliK}. The fitted values of the $q$ exponents are also
given in Tab.~\ref{tab:mft+log}. We should comment that logarithmic 
corrections are hard to isolate numerically, and that results are 
sensitive to assumptions about the size of the scaling region 
about the fixed point~\cite{HK}. Nonetheless, the values found do
suggest that the `GN' and `Thirring' formulations have 
qualitatively different scaling violations, the former resembling
the large-$N_f$ predictions and the latter the sigma model. How
stable the quoted $q$'s are, and whether the lattice-regularised
models fall into distinct universality classes (the notion is
still valid, since corrections to scaling can also be
universal~\cite{Zinn2}), are issues for further study.

Further support for the distinct nature of models with composite
scalars comes from a study of finite volume corrections when the
`GN' simulation is run directly in the chiral limit $m=0$~\cite{HK}.
On a finite system in this limit the order parameter $\Sigma$ is
strictly zero; we must instead monitor the quantity $\vert\Phi\vert$
defined by
\begin{equation}
\vert\Phi\vert=\biggl\langle\sqrt{{1\over2}\mbox{tr}\bar\Phi^\dagger
\bar\Phi}\biggr\rangle\;\;\;;\;\;\;\bar\Phi={1\over V}\sum_x
\left(\sigma(x)+i\vec\tau.\vec\pi(x)\right),
\end{equation}
where $\sigma,\vec\pi$ are the auxiliary fields. In Fig.~14 we plot
$\Delta\equiv\vert\Phi\vert-\Sigma_0$ versus $1/g^2$, where 
$\Sigma_0$ is the order parameter from the EOS (\ref{eq:eosmft})
extrapolated to the chiral limit. For a theory with fundamental 
  \begin{figure}
\setcounter{figure}{13}
  \centerline{
  \psfig{figure=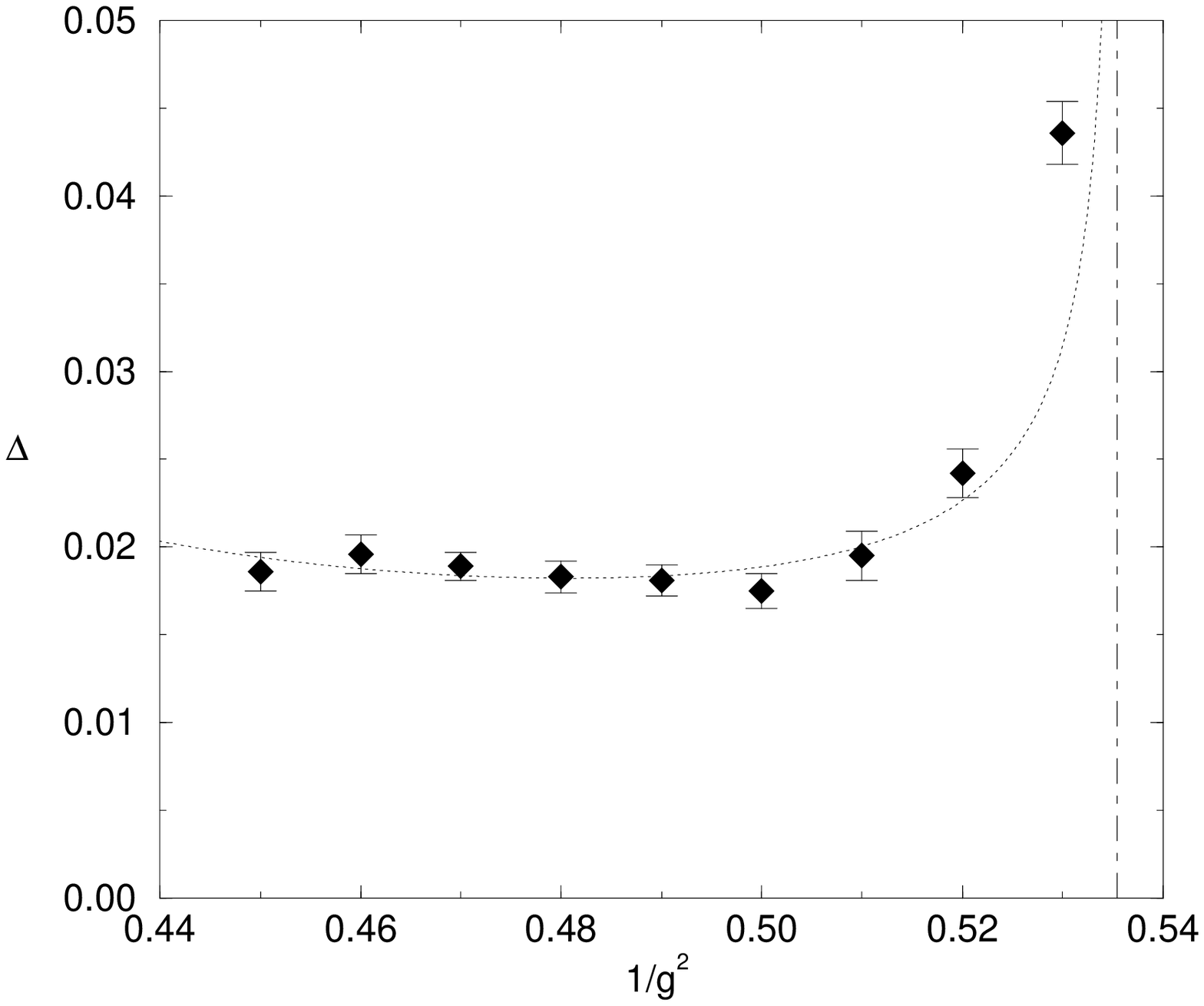,width=3.5in}
  }
\caption{Finite volume correction for the NJL model. The vertical
line shows the estimate for $1/g^2_c$.}
  \end{figure}
scalars, the correction is predicted to be of the form~\cite{Gock2}
\begin{equation}
\Delta\propto{1\over{L^2 \Sigma_0}},
\end{equation}
where $L$ is the linear size of the system ($L=16$ in this case).
For a composite scalar, however, in the large-$N_f$ approximation
the prediction becomes
\begin{equation}
\Delta\propto{1\over{L^2 \Sigma_0\vert\ln\Sigma_0\vert }}.
\label{eq:Delta}
\end{equation}
As shown in the figure, (\ref{eq:Delta}) provides a much more
convincing fit to the data.

\section{Conclusions}

To conclude, let me briefly summarise what each of the models we have looked
at has taught us:

\begin{enumerate}

\item[$\bullet$] The Gross-Neveu models with discrete 
and continuous chiral symmetries in $d<4$ appear under good
analytical and numerical control. They provide a paradigm for a
non-perturbative fixed-point theory containing elementary 
fermion and composite scalar
degrees of freedom, and a non-vanishing interaction at short
distances. Two further applications of these models, which I have
not had time to discuss, is that firstly unlike gauge theories they permit 
Monte Carlo simulation at non-vanishing chemical potential, and
hence non-vanishing baryon number density. This has been applied
in $d=3$ as a model effective theory of hadronic physics~\cite
{thermo}. In $d=4$ simulations of QCD~\cite{DKS} directly in the chiral limit
are possible by including
a gauge-invariant four-fermi interaction in the bare Lagrangian;
the resulting model lies in the same
universality class and hence contains the same IR physics as the true
theory~\cite{Brower}.

\item[$\bullet$] The Thirring model in $d<4$ has a fixed point for
small $N_f$ not described by the $1/N_f$ expansion. Genuine
non-perturbative techniques need to be applied, and the model deserves
to be used as a testbed for numerical lattice studies. The issue of
its equivalence with $\mbox{QED}_3$ is intriguing and needs to be
resolved.

\item[$\bullet$] There is lots more to learn about the nature of
triviality in $d=4$. So far in looking at corrections to the equation
of state we have pursued rather a ``condensed matter'' approach, and
argued that composite models have a distinct form of scaling violation
as compared to the standard ferromagnetic picture. It would be interesting
to try to extract particle physics from these models, for instance, by asking
whether they predict new triviality bounds on the Standard Model
Higgs mass.
Finally, there remains the fascinating possibility of
actually finding new non-trivial fixed points in $d=4$ by expanding
the space of couplings, say by introducing a gauge interaction such as
in strongly-coupled QED~\cite{QED4}.

\end{enumerate}

  \section*{Acknowledgments}
  My work is supported by a PPARC Advanced Research 
  Fellowship. It is a pleasure to thank Luigi Del Debbio, Sasha Koci\'c
and John Kogut for enjoyable collaboration,
and my Korean hosts for their excellent hospitality.

  \end{document}